\documentstyle[appb,epsf]{article}
\def\bei{\begin{itemize}}
\def\eei{\end{itemize}}
\def\ben{\begin{enumerate}}
\def\een{\end{enumerate}}
\def\l{\label}

\def\mp{m_\pi}
\def\y{\frac{e^{-\mp r}}{{ r}}}
\def\v#1{\vec{#1}}
\def\b#1{\bar{#1}}
\def\h#1{\hat{#1}}

\def\Ms{M^*}

\def\vPpa{\frac12\v P+\v p}
\def\vPpb{\frac12\v P-\v p}
\def\vPppa{\frac12\v P+\v p\,'}
\def\vPppb{\frac12\v P-\v p\,'}

\def\n{\nonumber \\}
\def\e{&=&}

\def\tt{\v \tau_1\cdot\v \tau_2}

\def\gmsq{(\frac{g_{\pi NN}m_\pi}{2M})^2}

\def\lNM{\langle {\rm N \,Matter}\mid}
\def\rNM{\mid  {\rm N \,Matter}\rangle}

\begin{document}
\title{\hfill DOE/ER/40762-150 \\
\hfill U.Md. PP\# 98-125 \\
\hfill\mbox{}\\
\hfill\mbox{}\\
Relativity Damps OPEP in Nuclear Matter\thanks{In honor of Joseph Speth's 60th birthday}} \author{Manoj
K. Banerjee
\address{ Department of Physics, University of Maryland, \\
College Park, MD, USA, 20742-4111\\
e-mail: banerjee@quark.umd.edu\\}}

\maketitle

\date{\today}

\begin{abstract}
Using a relativistic Dirac-Brueckner analysis the OPEP 
contribution to the ground state energy of nuclear matter is
studied.  In the study the pion is derivative-coupled. We find that the role of  the tensor force 
in the saturation mechanism is
substantially reduced compared to its dominant role in a usual 
nonrelativistic treatment. 
We   show that the damping of derivative-coupled OPEP is actually due to the 
decrease of $M^*/M$ with increasing density.  We point out that if derivative-coupled OPEP is the preferred form of nuclear effective lagrangian nonrelativistic treatment of nuclear matter is in   trouble.  Lacking the notion of $M^*$  it cannot replicate the damping. We suggest an examination of the feasibility of using  pseudoscalar coupled $\pi$N interaction before reaching a final  conclusion about nonrelativistic treatment of nuclear matter.

\end{abstract}

The talk is based on a recent paper written by Tjon and myself~\cite{MKBTJ}.

The one pion exchange potential, OPEP, in
  momentum space is given by the expression:
\begin{eqnarray}&&\langle \vPppa,\vPppb\mid v\mid \vPpa,\vPpb\rangle= \n
&&-\gmsq\,\frac{\v \sigma_1\cdot(\v p\,'-\v p) \v \sigma_2\cdot(\v p\,'-\v p)}{(\v p\,'-\v p)^2+\mp^2}\v\tau_1\cdot\v \tau_2.\nonumber\end{eqnarray}

With $\frac{1}{4\pi}\gmsq\,\simeq 11{\rm MeV}$ the OPEP in coordinate space is given by the expression:
\begin{eqnarray}
v_\pi(\v r)\e 11 Mev\left\{-4\pi\frac{\v\tau_1\cdot\v \tau_2}{3}\,\v \sigma_1\cdot\v \sigma_2\,\frac{\delta(\v r)}{\mp^2}\right.\n
&+&\frac{\v\tau_1\cdot\v \tau_2}{3}\,\v \sigma_1\cdot\v \sigma_2\,\y\n
&+&\left.\frac{\v\tau_1\cdot\v \tau_2}{3}\,S_{12}\,\y(1+\frac{3}{\mp r}+\frac{3}{\mp^2 r^2})\right\}\nonumber\end{eqnarray}
where the last term is the tensor force and $$S_{12}= 3\v \sigma_1\cdot\h r\v \sigma_2\cdot\h r- \v \sigma_1\cdot\v \sigma_2$$  is the tensor operator.  In a Hartree-Fock calculation in nuclear matter the $\delta(\v r)$ potential is capable of contributing at saturation density, $\rho_0\simeq  \frac12 m_\pi^3$,  as much as $\sim 20$MeV. But we must ignore it as it will be wiped out by the short-range correlation. The Yukawa potential and the tensor force contribute  a mere $\simeq \,-2$MeV.  Really important  contributions of the tensor force come from second and higher orders. The large matrix element, $\langle ^3D_1\mid S_{12}\mid^3S_1\rangle =\sqrt{8}$, shown in the matrix below,
\vspace{0.25in}
$$S_{12}^2=8-2S_{12}.$$
\begin{center}
\begin{tabular}{|c|c|c|}\hline
&$ \mid^3S_1\rangle$&$\mid ^3D_1\rangle$\\ \hline
$\langle ^3S_1\mid$&$0$&$\sqrt{8}$\\ \hline
$\langle ^3D_1\mid$&$\sqrt{8}$&$-2$\\ \hline
\end{tabular}
\end{center}
 is responsible for this feature.

The effect of the tensor force and its dominance in  nonrelativistic nuclear physics are seen most dramatically from the following results for the deuteron~\cite{WSS}.
$$\langle Deuteron\mid V_{\rm central}\mid Deuteron\rangle \sim 0$$
$$2\langle ^3D_1\mid V_{\rm tensor}\mid ^3S_1\rangle \sim -22 {\rm MeV}$$
$$$$
 $$\langle Deuteron\mid \frac{\v p^2}{2M}\mid Deuteron\rangle \sim +20 {\rm MeV}$$

In a nonrelativistic Bethe-Brueckner  calculation of nuclear matter one finds
typically~\cite{FG}
\begin{equation}
\lNM V_\pi\rNM_{\rm {nonrelativistic}} \sim -34\,
(\rho/\rho_0)^{0.45}\,{\rm MeV}. 
\l{vpinr}
\end{equation}  
The exponent of the density $\rho$ is  markedly less than the nominally 
expected value of $1$ because of Pauli blocking.

Relativistic nuclear physics is heavily based on summing Bethe-Salpeter ladders using some form of Blankenbecler-Sugar-Logunov - Tavkhelidze~\cite{BSu}  prescription to obtain quasipotentials. In general quasipotentials are not simply described by tree graphs with dressed vertices.  The OPEP is an exception. Other terms have the characteristics of four-point functions. In practice, one uses OBEP forms with form factors for approximate representation of  quasipotentials. The parameters are fixed by fitting NN data.

The relativistic results for the contribution  of $V_\pi$ to the
deuteron~\cite{HumTj} is
$$\langle {\rm Deuteron}\mid V_\pi\mid {\rm  Deuteron}\rangle_{\rm relativistic} =-22{\rm MeV},$$
suggesting an equally important role of the pion. In sharp contrast, 
this seems not to be the case in a relativistic treatment of nuclear matter.

Strong scalar ($S$) and vector ($V$) fields
of the order of a few hundred MeV are typical for  relativistic
theories~\cite{mth,AmTj,Mach} based on a meson theoretical
description of the nuclear force. These values are
consistent with expectations based on the studies of scattering of $\sim
1$ GeV protons by nuclei. The large scalar fields have far reaching
consequences in nuclear matter through the strongly medium
modified nucleon mass $M^*=M+S$. The saturation mechanism is believed to
rest upon the decrease of magnitude of $S$ with increasing
density.  Of course,  in a
mean field theory (MFT) like the QHD~\cite{SW} it is the only possible mechanism for saturation.

The  contributions
of a particular meson exchange potential, $V_\alpha$,  to the binding energy can be calculated  using the
Hellmann-Feynman theorem
\begin{equation}
\lNM V_{\alpha}\rNM=g_{\alpha NN}^2\frac{\partial}{\partial \,g_{\alpha NN}^2}(E/A).
\l{Vpi} \\
\end{equation}
We  find the following for the pion field contribution to $E/A$:    
\begin{equation}
\lNM
V_\pi\rNM_{\rm Relativistic} \sim -20\,(\rho/\rho_0)^{0.16}\,{\rm MeV}.
\l {ROPEP}
\end{equation}
>From this we see  that the pion contribution is considerably suppressed  
compared to the  
value given by Eq.~(\ref{vpinr}) for the nonrelativistic case. 
We will make clear that the suppression of OPEP 
is generic and not particular to the present calculation. 
Furthermore, OPEP has only a minor role in the saturation mechanism.  
This is exhibited in Fig.~\ref{novpi} where we plot 
our calculated results of $E/A$ (curve a)
and $E/A-\langle NM\mid v_\pi\mid NM\rangle -17\,$MeV (curve b).
The two curves have practically the same density dependence verifying that OPEP
contributes little to the saturation mechanism. The subtraction of $17$ MeV in
curve (b) makes the scale more compact.
\begin{figure}[htb]

\vspace{0.5in}

\epsfxsize=2.2in
\epsfysize=2.2in
%\vspace{0.2in}
\hspace{1in}\epsffile{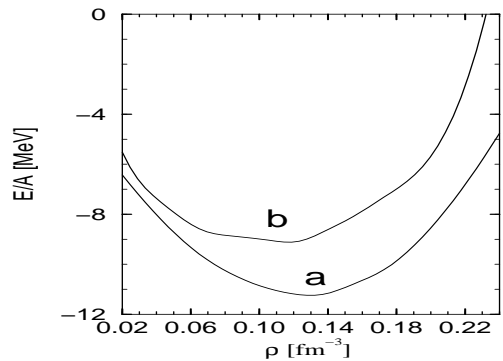}

  \caption[F4]{Plots of the Dirac-Brueckner predictions of $E/A$ (curve a) and
   $E/A$$-$\mbox{$\langle NM\mid v_\pi\mid NM\rangle$}$-$$17\,$MeV (curve b).
              }
%The subtraction of $17$MeV in curve (b) makes the scale more compact.}
\l {novpi}
\end{figure}

The preceding numbers confirm the belief that in a relativistic treatment of nuclear matter the tensor force does not have the dominant role that it has in the usual nonrelativistic treatment. The  radically different explanations of the saturation mechanism in 
nonrelativistic and relativistic studies   of nuclear matter 
constitute  a puzzling issue. A valid nonrelativistic treatment 
must reproduce the main physics of  a valid relativistic treatment 
in leading order in $v/c$. 
Although the issue is a longstanding one,
%already exists since the early days of relativistic nuclear physics, 
no resolution of it has been given  to date. We have addressed this question.
A Dirac-Brueckner (D-B) analysis~\cite{mth,AmTj} is at present
the best tool we have for a relativistic study of nuclear matter.  
We examine here the role of derivative-coupled OPEP in D-B and
show that it is substantially reduced due to relativity.  Since the 
contribution of OPEP to the deuteron binding energy remains large 
in a relativistic treatment the  damping in nuclear matter must 
be due to many-body effects.  We find that it can  be attributed 
to the decrease of $M^*/M$ with increasing density.

The above results can  be understood qualitatively by examining   the  second
order contributions to the $G$-matrix.
Keeping only the positive energy $\Ms$ state contributions in the
intermediate states we have, 
\newpage
\begin{eqnarray}
\langle \v p'\mid   G(P) && \mid  \v p\rangle = \langle \v
p'\mid  V(P)\mid\v p\rangle +\sum_{\lambda,i}\int\frac{d^3k}{(2\pi)^3}
\nonumber\\
&&
\times \langle \v p'\mid V(P)\mid\v k\rangle\frac{1}{M^*}\frac{  Q_{Pauli}} { 
(\v p/M^*)^2-(\v k/M^*)^2+\Delta/M^*}\langle \v k\mid  V(P) \mid\v
p\rangle.
\l{BGE3}
\end{eqnarray}

The tensor force contributes mainly to the second term  of Eq.~(\ref{BGE3}) which makes the structure of the two-nucleon propagator important.  Normally the boson masses in OBEP provide the scales for momenta in nuclear physics. But here we notice that the two-nucleon propagator  provides a new scale, viz., $M^*$. To exploit this new scale let us introduce dimensionless momenta, $\v \ell=\v p/M^*,\,\,\,\v n=\v k/M^*,\,\,{\rm etc.}$,  and exhibit a few OBEP matrix elements  in terms of these.
\begin{flushleft}
$\sigma$ Exchange
\end{flushleft}
\begin{eqnarray}
\langle \v p\mid v_\sigma\mid \v k\rangle \e -\frac{g_\sigma^2}{(\v p - \v k)^2+m_\sigma^2}\n
&=&-\frac{1}{M^{*\,2}}\frac{g_\sigma^2}{(\v \ell - \v n)^2+(m_\sigma/M^*)^2}. \label{vsigf}\end{eqnarray}
\newpage
\begin{flushleft}
$\pi$ Exchange (Derivative Coupling)
\end{flushleft}
\begin{eqnarray}
\langle \v p\mid v_\pi^{dc}\mid \v k\rangle \e (\frac{g_\pi}{2M})^2\v \tau_1\cdot\v \tau_2\frac{\v \sigma_1\cdot (\v k-\v p)\v \sigma_2\cdot (\v k-\v p)}{(\v p - \v k)^2+m_\pi^2}\n
&=&(\frac{M^*}{M})^2\frac{1}{M^{*\,2}}(\frac{g_\pi}{2})^2\v \tau_1\cdot\v \tau_2\frac{\v \sigma_1\cdot (\v  \ell-\v  n)\v \sigma_2\cdot (\v  \ell-\v  n)}{(\v \ell - \v n)^2+(m_\pi/M^*)^2}. \label{vpif}\end{eqnarray}

\begin{flushleft}
$\pi$ Exchange (Pseudoscalar Coupling)
\end{flushleft}
\begin{eqnarray}
\langle \v p\mid v_\pi^{\gamma_5}\mid \v k\rangle \e (\frac{g_\pi}{2M^*})^2\v \tau_1\cdot\v \tau_2\frac{\v \sigma_1\cdot (\v k-\v p)\v \sigma_2\cdot (\v k-\v p)}{(\v p - \v k)^2+m_\pi^2}\n
&=&\frac{1}{M^{*\,2}}(\frac{g_\pi}{2})^2\v \tau_1\cdot\v \tau_2\frac{\v \sigma_1\cdot (\v  \ell-\v  n)\v \sigma_2\cdot (\v  \ell-\v  n)}{(\v \ell - \v n)^2+(m_\pi/M^*)^2}. \label{vpi5f}\end{eqnarray}

Notice the differences in the $M^*$ factors. The derivative-coupled pion-exchange potential has an extra damping factor of $(M^*/M)^2$ relative to the sigma-exchange potential. It is reasonable to expect that the $M^*/M$ factor suppresses derivative-coupled OPEP, Pauli coupled $\rho$, etc. 

No such damping factor is present for the pseudoscalar coupled pion-exchange potential. To our knowledge all published relativistic nuclear matter calculations have used derivative-coupling. Of course, the reason is well-known. The pair suppression problem is automatically taken care of with use of 
derivative-coupling. We will proceed as if derivative-coupling is correct. A discussion of pseudoscalar coupling vs. derivative-coupling follows at the end of the talk.
\begin{table}
\caption{Table of $M^*$ factors for various OBEP potentials. The number of stars in the 2nd column indicates the importance of the OBEP in nuclear interaction.}
\begin{center}
\begin{tabular}{|l|l|l|l|}\hline
Parity&Importance&Boson&Scaling Factor \\ \hline
E&* * *&$\sigma$&$(1/M^*)^2$ \\ \hline
V&* * *&$\omega$ Dirac&$(1/M^*)^2$ \\ \hline
E&&$\rho$ Dirac&$(1/M^*)^2$ \\ \hline
N&&$\delta$&$(1/M^*)^2$ \\ \hline
O&* *&$\pi$ Der. Coupled&$(M^*/M)^2(1/M^*)^2$ \\ \hline
D&*&$\rho$ Pauli&$(M^*/M)^2(1/M^*)^2$ \\ \hline
D&&$\eta $ Der. Coupled&$(M^*/M)^2(1/M^*)^2$ \\ \hline
&-*&$\omega$ Pauli&$(M^*/M)^2(1/M^*)^2$ \\ \hline
\end{tabular}
\end{center}
\end{table}

The $M^*/M$ suppression is corroborated in more detail by
the following calculation.  Let us modify the $S$ obtained from 
the self-consistent D-B calculation by multiplying it with the factor 
$\alpha\leq 1$ thus generating a $M^*=M+\alpha S$.  By  using the
modified scalar self-energy in the nucleon propagators we recalculate 
first the $G$ matrices and then $E/A$ and finally $\lNM V_{\pi}\rNM$ 
using Eq.~(\ref{Vpi}). Only the $\alpha=1$ analysis is self consistent;
others are not. But such a calculation is particularly suitable to
exhibit the role of $M^*/M$ on the OPEP contribution.
Figure~\ref{vpisoft} exhibits clearly the damping  due to 
decreasing $M^*/M$.
We stress that the mechanism of damping is generic to any relativistic 
treatment using derivative-coupled pion and not particular to either Ref.~\cite{AmTj} or 
the use of Ref.~\cite{BSu}.
\begin{figure}[htb]

\vspace{0.5in}
\epsfxsize=2.2in
\epsfysize=2.2in
%\vspace{0.2in}
\hspace{1in}\epsffile{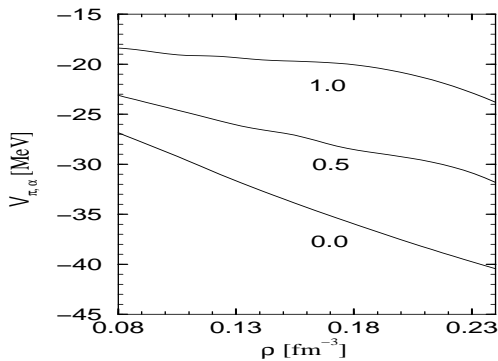}
\vspace{0.2in}
\caption[F4]{Plots of $\langle NM\mid V_\pi\mid NM\rangle$ with the parameters of
Ref.~\protect{\cite{AmTj}}. In the G-matrix calculations $S$ is replaced with
$\alpha S$. The plots are for $\alpha=0.$, $0.5$ and $1.0$. The last one is the
result of  a D-B self-consistent calculation. The
other two are not  self consistent.}
\l {vpisoft}
\end{figure}

We want to be careful that the present work not be interpreted as
providing support for MFT.  Results of calculations of $E/A$  using the same interaction, namely, that of Ref.~\cite{AmTj}, in both D-B and MFT treatments are shown in  Fig.~\ref{mfte}.  We see that the results are distinctly different. Such differences are found in the results for scalar and vector fields in the two treatments. Results obtained upon using the intearction of Ref.~\cite{AmTj} are listed below.
\begin{eqnarray}
S_{D-B}\e-306\,(\rho/\rho_0)^{0.81}\,{\rm MeV},\n
V_{D-B}\e 233\,(\rho/\rho_0)^{0.97}\,{\rm MeV}, \n
S_{MFT}\e - 358 \,(\rho/\rho_0)^{0.92}\, {\rm MeV},\n
V_{MFT}\e 295\,(\rho/\rho_0)\,{\rm MeV}\nonumber \end{eqnarray}

 Undoubtedly, if one releases oneself from the constraint
of fitting NN data and  freely chooses the NN interaction one can obtain
proper binding and saturation of nuclear matter with a MFT calculation.

\begin{figure}[htb]

\vspace{0.5in}
\epsfxsize=2.2in
\epsfysize=2.2in
%\vspace{0.2in}
\hspace{1in}\epsffile{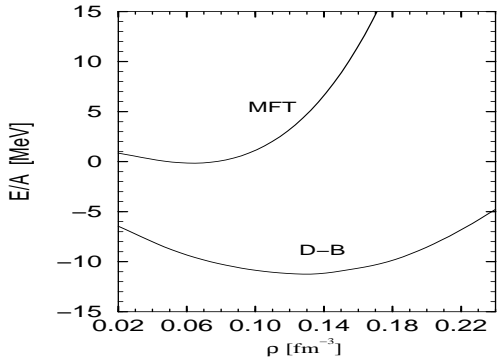}
%\vspace{0.2in}
\caption[F4]{Plots of $E/A$ from a  Dirac-Brueckner  
and a MFT calculation with the same quasipotential. }
\l {mfte}
\end{figure}
\begin{table}[tbp]
\caption{Amorim-Tjon and Bonn C parameters for derivative-coupling and standard pseudoscalar coupling parameters. The value of $m_\sigma$ is not specified.}
\begin{center}
\begin{tabular}{|l|l|l|l|l|l|l|}\hline
Coupling&Source&Isospin&$m_\sigma$&$g_\sigma$&$m_\pi$&$g_\pi$\\
&&&in MeV&&in MeV&\\ \hline
Derivative&Amorim-Tjon~\cite{AmTj}&&570&7.6&138&14.2\\ \hline
&Bonn C~\cite{Mach}&1&550&8.6&138&14.2\\ 
&&0&720&17.6&138&14.2\\ \hline
Pseudoscalar&&&?&?&138&14.2\\ \hline
\end{tabular}
\end{center}
\label{par}
\end{table}
The results presented in this talk are the first
explicit calculations showing that
in the relativistic treatment  the  tensor force
contributions generated by derivative-coupled OPEP\footnote{  Pauli $\rho$ tensor force is also damped.} are reduced in size in nuclear matter.  Because of this, in
complete contrast to the nonrelativistic situation, they cease to play 
an essential role in the saturation mechanism.
The reduction of the tensor force contributions is principally due to the
relativistic $\Ms/M$ effect.
But even the reduced role of OPEP is not negligible in the actual
saturation properties of nuclear matter. As noted,
it contributes $-20$MeV to $E/A$.
 The dominant mechanism of
saturation of nuclear matter is basically very different in the two approaches. In the
nonrelativistic approach it is the density-dependent reduction 
due to Pauli blocking of the attraction from tensor force,  
while in the relativistic approach it is the reduction of the rate 
of growth with increasing $\rho$ of the attraction from the  scalar  
field   relative to the  growth of  repulsion from the vector field. 

Finally, let us discuss the issue of derivative  versus pseudoscalar $\pi$N coupling.  If the former is the correct coupling for nuclear effective lagrangian then nonrelativistic treatment of nuclear matter appears not to be valid. On the other hand, if  a pseudoscalar coupling lagrangian can be found which gives satisfactory results for nuclear matter the nonrelativistic treatment may be valid. Unfortunately there are no published results for nuclear matter with  a pseudoscalar coupling lagrangian. Needless to say,  before doing any nuclear matter calculation the parameters  must be fixed by fitting NN data.

It is useful to remind ourselves that we have been dealing with quasipotentials.  These are still constrained to be chiral invariant. The   interaction lagrangians with which one could reproduce the quasipotentials via tree graphs with form factors are listed below.
\begin{eqnarray}
{\cal L}^{\rm derivative\,\, coupling}_{\pi\,N}\e\frac{g_\pi}{2M}\b \psi \gamma_5\gamma^\mu\v \tau\psi\cdot \partial_\mu \v \pi +g_\sigma\b \psi\sigma\psi. \label{derlag}\\
{\cal L}^{ \gamma_5\,\,\rm  coupling}_{\pi\,N}\e g_\pi\b \psi[\sigma+i\gamma_5\v \tau\cdot\v \pi]\psi.
\label{gamma5lag}\end{eqnarray}

In the derivative-coupling lagrangian the nucleon fields are unaffected by chiral transformation and the scalar field $\sigma$ is a chiral singlet. In the pseudoscalar coupling lagrangian the nucleon fields, $\psi$ and $\b \psi$, belong to chiral $(1/2,0),(0,1/2)$ representations, while $\sigma$ and $\v \pi$ form  chiral $(1/2,1/2)$ representations.  In  Table \ref{par} we list the  parameters for two derivative-coupling lagrangians, namely, Amorim-Tjon~\cite{AmTj} and Bonn C~\cite{Mach}. The last line gives the standard pseudoscalar coupling lagrangian parameters. Notice that both the mass and the coupling constant of the $\sigma$ meson have been left unspecified. The reason is that, quite unlike OPEP, there will be considerable modification of the one-$\sigma$ exchange potential  as one goes from the form given by the original lagrangian to the quasipotential. This happens  through two distinct mechanisms.  First,  the $\sigma$ couples to the pion clouds of each of the pai!
r of interacting nucleons. Second,
 a pair of  interacting  pions are exchanged between two nucleons. The interaction must be isovector in the two-pion t-channnel\footnote{For example, by exchanging a $\rho$ meson.}. The prospect of a pseudoscalar lagrangian succeeding in the nuclear matter problem is not very good. The undamped tensor force will contribute an additional $\sim -15\,$MeV. To compensate this the effective $\sigma$ nucleon coupling in the quasipotential must decrease. It is difficult to see how such a reduction can come about.  Still, the only recourse is to actually carry out the program of study with a pseudoscalar lagrangian before we come to a definitive conclusion about the future nonrelativistic treatment of nuclear matter.

This research is supported by the U.S. Dept. of Energy under grant no. DE-FG02-93ER-40762.


\begin{thebibliography}{100}
\bibitem{FG} This result was generated with a Brueckner theory program of Gruemmer (J\"{u}lich) using the Bonn C potential described in Ref.~\cite{Mach}.
\end{thebibliography}
\end{document}